\documentstyle[prd,preprint,aps]{revtex}

\newcommand{\be}{\begin{equation}}
\newcommand{\ee}{\end{equation}}
\newcommand{\bea}{\begin{eqnarray}}
\newcommand{\eea}{\end{eqnarray}}

\newcommand{\V} {{\cal V}}

\begin{document}

\reversemarginpar
\tighten

\title{Of Bounces, Branes and Bounds} 

\author {A.J.M. Medved}

\address{
Department of Physics and Theoretical Physics Institute\\
University of Alberta\\
Edmonton, Canada T6G-2J1\\
E-Mail: amedved@phys.ualberta.ca\\}

\maketitle

\begin{abstract}

Some recent studies have considered  a  Randall-Sundrum-like
 brane world evolving in the background of an anti-de Sitter
Reissner-Nordstrom black hole.  
For this scenario, it has been
shown
that, when the bulk charge is non-vanishing, 
a singularity-free ``bounce'' universe
will always be obtained. However, for  the physically relevant  case of a  
de Sitter brane world, we have recently argued
that, from a holographic ($c$-theorem)
perspective, such brane worlds may not be physically
viable. In the current paper, we reconsider
the validity of such models by appealing to
the so-called ``causal entropy bound''.  In this framework,
a paradoxical outcome is obtained:
these brane worlds are  indeed holographically  viable, 
provided that the bulk  charge 
is not too small. We go on to argue that this new finding
is likely the more reliable one.

\end{abstract}

\section{Introduction}

\par

Progress in understanding gravity, whether at a 
classical, semi-classical or quantum level, often requires a
well-defined notion of what constitutes physically realistic matter.
With guidance from the observable universe, most physicists
would agree on matter that has a positive energy density
(or, at least, a limit on the degree of negativity)
and that preserves causality ({\it i.e.}, signals should not
exceed the speed of light). Historically speaking,
 these notions were put on a rigorous footing
by the well-known  energy conditions of  general relativity
\cite{WALD}.  From a modern perspective, the most
prominently called upon of these constraints is
the {\it null energy condition}, which can
be expressed for perfect-fluid matter as follows:
\be
\rho+p\geq 0,
\label{1.1}
\ee
where $\rho$ is the energy density and $p$ is
the pressure. Since causality, by itself, further implies that 
$|p|\leq|\rho|$\footnote{There are, however, subtle examples in which
causality can be maintained even if $|p|\leq |\rho|$ is violated.
See the introduction of \cite{MCI} for an interesting discussion.}, the null
energy condition really  suggests
that $\rho \geq 0$.  (The only possible exception
is when $\rho=-p < 0$, which is the well-understood case
of a purely anti-de Sitter spacetime.)
\par
At the classical level, the null energy condition holds up
relatively well and  has, at times, been elevated
to almost the status of a fundamental principle.
However, classical violations are known to occur;
especially in theories that incorporate non-minimally
coupled scalar fields, such as inflationary cosmology.
At the quantum level, the situation is even worse,
 as violations of the null condition due to quantum effects
 are certainly not an uncommon event.
(For recent discussions and relevant citations, see \cite{BV,MCI2}.)
\par
In spite of its somewhat dubious status, the null energy condition
continues to be utilized in many phases of theoretical
gravity research.\footnote{Sometimes the {\it weak energy
condition} \cite{WALD} is alternatively employed, which is,
with the assumption of causality, essentially
the  same as the null energy condition.}
The motivating factor for this persistence is
that the violations which do occur seem to be relatively small,
implying that some sort of bound on negative energy
densities must still be in effect. 
Unfortunately, what this bound should  be, 
precisely,  remains conspicuously unclear. (Averaging the 
energy densities  over space and/or
time is, however, one distinct possibility  \cite{FORD}.)
\par
If we accept that spacetimes with negative-energy matter 
can, in principle, exist (thereby  rendering the null energy 
condition as obsolete), then  a method of
discriminating between  the plausible and the implausible
becomes a critically important issue. 
One  possible recourse is to appeal to the holographic principle,
which is believed to be a fundamental element of
any viable theory of quantum gravity \cite{THO,SUS,BOU2}.
In essence, the holographic principle provides
a bound on the amount of information or entropy that
can be stored in a given region of spacetime.
 It thus follows that a clear violation
of a suitably defined holographic bound  - or some other 
manifestation of this paradigm, such as
a holographically induced $c$-theorem \cite{MAL} -
should provide ample evidence that a given spacetime
is physically unacceptable.\footnote{Holographic
 bounds, however, can  take  on substantially different forms
depending on the context. Hence, the
 applicability of holography in a given situation is not always 
straightforward. We discuss this point
further in Section 2.}
\par
There is a caveat that  should be considered before
one attempts to apply the holographic principle
in the above manner. Namely, many formulations
of this  principle have used the null (or weak)
energy condition as an  antecedent. For instance, let us consider 
the {\it covariant holographic bound} as proposed by Bousso
\cite{BOU}. Although this bound is believed 
to  have universal validity \cite{BOU2}, it has only been rigorously
established under certain conditions \cite{FMW};  one of these being the
null energy condition.  That is to say,
the covariant bound might require modifications (albeit, likely minor)
if it is to be applied to spacetimes with
a limited degree of negative-energy matter. The null (or weak)
energy condition also figures prominently in the 
formulation of holographically induced $c$-theorems \cite{MAL}.
\par
Notable examples of the  ``holographic discrimination''
of  {\it exotic} spacetimes\footnote{In this paper,
exotic will often be used as a synonym for the presence
of negative-energy matter.}
include recent works by  Brustein {\it et al} \cite{BFM}  
and McInnes \cite{MCI2}. What is of particular interest
is that both of these treatments have employed aspects of
holography which nicely circumvent the caveat discussed
above. More specifically, the former utilized the 
{\it causal entropy bound} of Brustein and Veneziano \cite{CAU}, while
 the latter
applied a holographic consistency condition that was
first proposed by Bousso and Randall \cite{BAR}. 
\par
In the current paper, our
 primary objective  is a similar type of holographic viability
test on a certain class of exotic cosmologies; in particular,
 we will follow \cite{BFM}  
in utilizing 
the causal entropy bound \cite{CAU}. 
As for the model  to be tested, it is essentially
a Friedmann-Robertson-Walker (FRW) universe  with radiative matter and dark
energy,
but  also containing {\it stiff} matter 
with a negative energy density ({\it i.e.}, matter for which 
$p=\rho<0$). Such cosmologies are of interest for
a couple of reasons. Firstly, they naturally arise in  Randall-Sundrum-like
brane-world scenarios \cite{RS} for which a three-brane ({\it i.e.}, 
``our universe'')  is moving
in the five-dimensional background of an  anti-de Sitter 
Reissner-Nordstrom black
hole.\footnote{Some other treatments of brane cosmology with
a bulk charge can be found in 
\cite{BC1,NEW,BC2,BC3,BC4,BC6,BC7,WOO,MP,AJM}.} 
In this regard, it is useful to keep in mind that,
from the perspective of a brane observer,
the motion of the brane through the (otherwise) static bulk
will appear as either a cosmological contraction
or expansion \cite{KRA}. Secondly, it has been
shown that the effective brane cosmology is that of
a ``bouncing'' FRW universe, completely devoid of
singularities \cite{MP,AJM}. Such bounce universes 
provide an intriguing means of circumventing
the theoretical complications of the ``big bang'' (or ``crunch'').
\par
To further motivate the upcoming analysis,
let us take note of a recent paper (by the current author) \cite{AJM}.
In this work, we subjected the very same brane-world-induced bounce
cosmologies to a holographic litmus test of a different kind.
The basis for this test (also see \cite{MCI,AJM3}) was a  
de Sitter-holographic
$c$-theorem \cite{DS1,DS2,DS3,DS4}, which can be viewed
as a consequence of the conjectured duality between  
de Sitter spacetimes and conformal field theories \cite{STR}.
(It is relevant that the brane worlds in question  are assumed
to have a positive vacuum energy and, hence, are typically
asymptotically de Sitter spacetimes.) Interestingly,
the results of \cite{AJM} imply that these cosmologies
are {\it not} physically viable from a holographic perspective.
However, because of the caveat discussed above, as well as some
interpretative difficulties (elaborated on in Section 5),
there is significant doubt as to the reliability of
the $c$-theorem in this particular context.
\par
The remainder of the paper is organized as follows.
In the next section, we discuss the causal entropy
bound \cite{CAU}; including its historical motivation,
the underlying premise and the relevant formalism.
In Section 3, we focus on the  brane-world scenario
of interest; specifically, the pertinent qualitative
and quantitative features of the induced FRW
cosmologies. However,  the formal derivations 
are left to the earlier works \cite{MP,AJM}.  In Section 4,
we  rigorously examine the holographic feasibility of
these spacetimes by way of the causal entropy bound.
Here, it is shown that these bounce cosmologies
are indeed viable, provided that the charge
(on the bulk black hole) satisfies a lower bound.
Finally, in Section 5,  we  provide a brief summary and
discussion; with particular emphasis  
on  interpreting  the current findings in view
of the paradoxical implications of our prior results \cite{AJM}.

\section{The Causal Entropy Bound}

\par
The causal entropy bound \cite{CAU} will
be central to the subsequent analysis,
so let us begin with a brief review
of its motivation and then  
present  the relevant formalism.
\par
The holographic principle \cite{THO,SUS} suggests that, for
a closed system, there will be some
finite upper bound on the entropy, $S$.
 On the basis of preserving the {\it generalized} second law of thermodynamics
\cite{BEK},
Bekenstein   proposed the following  bound
for a system of limited self-gravity:\footnote{For
the time being, we restrict considerations to
a four-dimensional spacetime and set all fundamental
constants, except for Newton's constant ($G_4$),
equal to unity. In this section, numerical factors (of the order unity)
 will often be ignored.}
\be
S < E R,
\label{2.1}
\ee
where $E$ is the total enclosed energy  and $R$ is 
a characteristic length scale (for instance, the
radius in the case of a spherical system).
The above can be rewritten as
\be
S<  {R R_S\over  G_4},
\label{2.2}
\ee
where $R_s\sim  G_4 E$
is the Schwarzschild radius of the system. Moreover,
because of the condition of limited gravity,
$R> R_S$ and the Bekenstein bound directly implies
\be
S< {R^2\over G_4} \sim {A\over G_4},
\label{2.3}
\ee
where $A\sim R^2$ is the surface area.
This is just the  ``usual'' form of the  holographic bound,
which follows from the viewpoint that
the entropy within a given volume should
be maximized by a  black hole of the same size \cite
{THO,SUS}.
\par
In  historical retrospect, it was not at all clear as
to how the above bounds should be extended  to scenarios of
strong or potentially strong gravity; in particular, cosmological
situations. Prior to the advent of holography, Bekenstein \cite{BEK2}
proposed that, for cosmological purposes,
$R$  in Eq.(\ref{2.1}) should be chosen as the particle horizon. 
This proposal, however,
does lead to contradictions, and so Fischler and Susskind \cite{FS}
suggested that Eq.(\ref{2.3}) should be modified so that
 the  area of the particle horizon 
 bounds the entropy contained in the past light cone. 
Although an improvement, the Fischler-Susskind bound
still runs into problems, such as during  the collapsing
phase of a closed universe.
Eventually, Bousso \cite{BOU} generalized their proposal
by considering the entropy bounded by  light cones
of diminishing cross-sectional area (which can be future and/or past directed
depending on the system in question). The Bousso bound has
the desirable features of being covariantly defined and having
general applicability ({\it i.e.}, not just cosmological scenarios),
and it has yet to be contradicted in any physically  realistic
situation \cite{BOU2}.
Nonetheless, the Bousso bound, which is formulated in terms
of null surfaces,  can {\it not} be universally   applied 
to spacelike regions in any given spacetime.  
\par
With regard to the issue of  holography in a spacelike
cosmological region, an interesting proposal
is the so-called  {\it Hubble entropy bound} \cite{VEN} (also
see \cite{EL}). This bound is based on a pair of
common-sense inputs: {\it (i)} the entropy
in a given region of space is maximized by
the largest black hole which can fit inside 
and {\it (ii)} in a cosmological background, the largest possible 
{\it stable} black hole has a radial size that  roughly corresponds
to   the Hubble 
horizon ($H^{-1}$).
By virtue of these observations, it immediately
follows that, in a given spacelike region, the  entropy 
should be bounded by the entropy of a Hubble-sized
black hole {\it times} the number of Hubble-sized
spheres that can  fit inside of  the total volume ($V$).
That is:
\be
S <  {H^{-2}\over G_4}\times {V\over H^{-3}}
= {VH\over G_4}.
\label{2.4}
\ee
\par
This  Hubble entropy bound served as the primary motivation
for the causal entropy bound of Brustein and Veneziano
\cite{CAU}. In fact, the causal bound can be viewed
as a covariant generalization of the Hubble bound.
More specifically, the Hubble horizon is replaced by a ``causal
connection scale'', $R_{CC}$,  that can be interpreted
as the length scale above which spacetime perturbations
are causally disconnected and, therefore,
black holes  can not (presumably) form. 
The determination of this length scale is a highly
technical process, and we refer the readers to
the seminal work \cite{CAU} for the details and
pertinent citations. Let us, however,
quote the result in a form that is readily
applicable to a standard (four-dimensional) FRW
spacetime (for the covariantly defined equivalent, again
see \cite{CAU}): 
\be
R_{CC}^{-2}={\rm Max}\left[{\dot H}+2H^2+{k\over a^2},\quad  -{\dot H}+
{k\over a^2}\right].
\label{2.5}
\ee
Here,
 a dot denotes differentiation
with respect to cosmological  time ($t$),
 $a=a(t)$ is the FRW scale factor,
$H={\dot a}/a$ is the Hubble ``constant'', and $k$
is the usual spatial-curvature parameter. Note
that $k$ takes on a value of 0, +1 or -1 for
a flat, closed or open universe (respectively).
\par
Actually, by assuming a perfect-fluid form 
for the spacetime matter  and applying
the well-known relations for the energy density ($\rho$)
and pressure ($p$) \cite{CAR},
\be
H^2={8\pi G_4\over 3}\rho -{k\over a^2},
\label{2.6}
\ee
\be
{\dot H}=-4\pi G_4 \left(\rho +p\right)+{k\over a^2},
\label{2.7}
\ee
one can translate Eq.(\ref{2.5}) into a
very convenient form:
\be
R_{CC}^{-2}=4\pi G_4{\rm Max}\left[{\rho\over 3}-p,\quad \rho+p\right].
\label{2.8}
\ee
It may be surprising that  this particular expression
has no explicit dependence on the curvature parameter, $k$.
\par
Let us now be more explicit with regard to
the proposed bound. 
The total entropy
contained in a spacelike  hypersurface of
volume V  should be bounded according to ({\it cf}, Eq.(\ref{2.4}))
\be
S< \beta {VR_{CC}^{-1}\over G_4},
\label{2.9}
\ee
where $\beta$ is a numerical factor  of the order
unity (reflecting the factors left out of Eq.(\ref{2.4})
and the inherent ambiguity in  bounds of
this nature).
Note that the causal connection scale can, for the problems of interest,
be  calculated by way of Eq.(\ref{2.5}) or (\ref{2.8}).
\par
Although the causal entropy bound has a conjectural
status, let us take note of the following
supporting evidence: {\it (i)} violations of
the bound require either trans-Planckian temperatures,
matter sources with a negative energy density or
an acausal equation of state \cite{BFM}
(all of which are arguably, but not necessarily, unphysical conditions),
{\it (ii)}  the bound closely follows Bousso's covariant
bound \cite{BOU} in situations where they can be  compared
\cite{CAU}, 
{\it (iii)} in fact, in its explicitly covariant
form \cite{CAU}, the causal bound is closely related
to a condition that was used by Flanagan {\it et al} \cite{FMW} to
derive the Bousso bound, 
{\it (iv)} the causal bound is parametrically
equivalent to the various holographic bounds
proposed by Verlinde \cite{VER} for a closed, radiation-dominated
universe \cite{BFV}, 
{\it (v)} for conventional cosmologies, the bound essentially
translates into $S<\sqrt{E V/G_4}$,  a form that has also
 turned up  in studies on uncertainty
relations \cite{SAS} and  extensive
thermodynamic systems \cite{TIM}.
\par
It is useful to keep in mind that, although the
causal bound works most favorably for
positive-energy matter, it can still persist
when some of the matter sources have a negative energy density \cite{BFM}.
That is to say, the causal bound can, 
in principle,  be used to test the validity of
exotic cosmological models
that might otherwise be rejected on the basis
of the null energy condition. 
It is this particular feature of the 
causal bound that makes it an ideal 
testing ground for the  brane cosmologies discussed below.

\section{Induced Brane-World Cosmologies}

Let us now  formally introduce the model of interest,
which can be viewed as a generalization of the Randall-Sundrum
brane-world scenario \cite{RS}. More specifically,
we will consider a 3+1-dimensional (positively curved\footnote{This
condition of positive curvature is for the sake of compliance
with the empirical evidence \cite{CAR} but is actually
inconsequential to the arguments which follow.}) brane 
moving in a 4+1-dimensional bulk spacetime
that is otherwise {\it static}. Without loss of generality,
the bulk geometry can be described by an anti-de Sitter
black hole with an electrostatic charge and a constant-curvature
horizon \cite{NEW}.  Such black holes are ``Reisnner-Nordstrom-like'',
but with an arbitrary horizon topology \cite{BIR}. 
\par
In a general sense, the above type of picture is known
to lead to a brane dynamical equation that mimics
a FRW universe \cite{KRA}. Moreover, the details
of the bulk solution play a prominent role in determining
what kinds  of  matter will be  {\it holographically induced}
on the brane world. (There should, of course, also be matter 
that lives strictly on the brane; most prominently, the
contributions from the standard model. This matter
can, however,  be ignored for the current, ``philosophical''
discussion.) For a charged black hole bulk in particular,
the brane world  turns out to be  a ``bounce'' cosmology
\cite{MP,AJM}.
That is, the universe is asymptotically de Sitter 
in the far past, contracts to a non-vanishing minimum at
some given time (which we take as being  $t=0$ without
loss of generality) and then expands to an
asymptotically de Sitter future.\footnote{Actually,
for the Reissner-Nordstrom  ($k=+1$) case, 
a sufficiently massive bulk black hole
may end up as an eternally oscillating universe
rather than   ``escaping''  to an asymptotically de Sitter
cosmology at temporal infinity.
This distinction has, however, no relevance
to our subsequent analysis.}
The important point to keep in mind is that
the bounce is strictly non-singular ({\it i.e.}, the FRW scale factor
does {\it not} vanish) provided that the bulk black hole
has a non-vanishing charge, no matter how small.
\par
A detailed discussion of this cosmological framework 
(at least for the asymptotic regimes) 
can be found in \cite{MP,AJM} (also see \cite{PS,AJM2}),
and we refer the interested reader to these citations.
Here, we will only quote the formalism that is
necessary for current purposes.
\par
Firstly, let us take note of the bulk solution;
namely, a five-dimensional anti-de Sitter Reissner-Nordstrom-like
black hole: 
\be
ds^2_{5}=-f(r)d\tau^2+{1\over f(r)}dr^2+r^2d\Omega^2_{k,3},
\label{3.1}
\ee
where
\be
f(r)= {r^2\over L^2}+1-{\omega M\over r^{2}}
+{3\omega^2 Q^2\over 16 r^4}
\label{3.2}
\ee
and
\be
\omega\equiv {16\pi G_5\over 3 \V_3}.
\label{3.3}
\ee
In the above,  $L$ is the curvature radius of the anti-de
Sitter bulk, 
$G_5$ is the five-dimensional Newton constant
(typically, $G_5\sim L G_4$), and 
$\V_3$ is the dimensionless
volume element associated with the three-dimensional (spacelike) 
constant-curvature hypersurface 
$d\Omega^2_{k,3}$.
There are also three constants of integration
for this solution: 
{\it (i)} the discrete parameter
$k$, which describes the horizon topology 
such that +1, 0 and -1 respectively correspond to
a spherical, flat and hyperbolic geometry, {\it (ii)}
the conserved mass of the black hole, $M$,  
which can be regarded as 
a strictly non-negative quantity\footnote{Technically speaking, the 
hyperbolic ($k=-1$) black hole solution
supports a negative mass; however, this {\it very} exotic scenario will
not be considered here.} and
{\it (iii)}  the electrostatic charge, $Q$, of the black hole.
For the duration, we will rather work with the
dimensionless measure of charge
\be
\epsilon^2 \equiv 3 Q^2 /4 M^2. 
\label{3.4}
\ee
It can be shown that 
the existence of a pair of positive and real
black hole horizons necessitates that $\epsilon^2<1$.
Moreover, on intuitive grounds, we will 
assume that $\epsilon$ is at least an order
of magnitude smaller than one.
\par 
Let us next focus our attention on
the brane world. After some suitable identifications,
one finds that the induced metric on the
brane takes on the following FRW form:
\be
ds^2_{4}=-dt^2+a^2(t)d\Omega^2_{k,3},
\label{3.5}
\ee
where $t$  measures the physical time
from the point of view of a brane observer and  
$a=r(t)$ is the cosmological   scale factor. 
 The corresponding Friedmann equation
 was  found to be as follows \cite{AJM}:
\be
H^2= {\Lambda_{4}\over 3}-{k\over a^2}+{\omega M\over a^4}
-{\omega^2 M^2 \epsilon^2\over 4 a^6},
\label{3.6}
\ee
where the Hubble constant, $H$, is as defined in
the prior section, and  $\Lambda_4$ is a constant that depends on
both $L$ and the tension of the brane. Obviously,  $\Lambda_4$ 
plays the role of the (effective) cosmological
constant in the brane universe.  Lacking a microscopic
theory, the best we can do is to fix $\Lambda_4$
according to the observational data \cite{CAR}; hence,
$\Lambda$ will be regarded as being a positive quantity with
an extremely small magnitude ($\sim 10^{-120}$ in Planck
units).
\par
Remarkably, Eq.(\ref{3.6}) is simply the four-dimensional
Friedmann equation for radiative matter (the $a^{-4}$ term)
along with an exotic (negative-energy) stiff-matter source,
the $a^{-6}$ term. 
Note that this brane universe
can be open, closed or flat ($k=-1$, +1 or 0) depending on the 
horizon topology of the bulk solution.
\par
To help clarify matters, we can suggestively rewrite the Friedmann
equation in the following manner:
\be
H^2 +{k\over a^2} ={8\pi G_4\over 3}\left[\rho_{v} + \rho_{r} +\rho_{e}\right],
\label{3.7}
\ee
where the {\it vacuum}, {\it radiation} and
{\it exotic} energy densities have respectively
been defined as follows:
\be
\rho_{v}={\Lambda_4\over 8\pi G_4},
\label{3.8}
\ee
\be
\rho_{r}={3M\over 8\pi G_4}a^{-4},
\label{3.9}
\ee
\be
\rho_{e}=-{3\omega^2 M^2 \epsilon^2\over 32 \pi G_4} a^{-6}.
\label{3.10}
\ee
The corresponding pressures are also known by way
of standard identifications \cite{CAR}:
\be
p_{v}=-\rho_v,
\label{3.11}
\ee
\be
p_{r}=\rho_r/3,
\label{3.12}
\ee
\be
p_{e}=\rho_e.
\label{3.13}
\ee
\par
As discussed earlier, one finds 
 \cite{MP,AJM} that the brane cosmology describes a bounce universe:
the scale factor reaches a finite minimum at $t=0$
and is (typically) asymptotically de Sitter in both the
distant past and far future. The bounce can be attributed
to the negative-energy matter, which dominates at
small values of $a$ (thanks to the $a^{-6}$ red-shift factor)
and creates a significant enough repulsive force so
that a ``big crunch'' is avoided. 
\par
Although
the scale factor can not be explicitly solved for
throughout its evolution (unless we set $\epsilon$ \cite{PS,AJM2}
or $\Lambda_4$ \cite{MP} to vanish), the asymptotic regimes
are well understood because of the discrepancy in
the various red-shift factors. More to the point, 
the constant vacuum energy plays essentially no
role near the bounce, whereas the other sources are
rapidly diluted away as $|t|$ or $a$ becomes large.
For the current analysis, we are particularly interested
in the solution near the bounce (see
the following section), which is,
up to negligible corrections, expressible as follows \cite{MP,AJM}:
\be
a^2={\omega M\over 2}\left(1-\sqrt{1-\epsilon^2}\cos\left[2\eta\right]
\right) \quad\quad\quad {\rm if}\quad k=+1,
\label{3.14}
\ee 
\be
a^2=  {\omega M\over 4}\left(\epsilon^2+4\eta^2\right)
\label{32.5} \quad\quad\quad {\rm if}\quad  k=0,
\label{3.15}
\ee
\be
a^2={\omega M\over 2}\left(\sqrt{1+\epsilon^2}\cosh\left[ 2\eta\right]
-1\right) 
\label{3.16} \quad\quad\quad {\rm if}\quad  k=-1,
\ee
where $\eta$ is the usual conformal-time coordinate
({\it i.e.}, $dt=a d\eta$).
\par
It is clear from the above forms that, as long as the charge
is non-vanishing ({\it i.e.}, as long as $\epsilon^2>0$),
$a$ will never shrink to zero.  At the bounce ($t=\eta=0$) in particular,
all three equations take the simple form
\be
a^2(0)={\omega M \epsilon^2\over 4} +{\cal O}[\epsilon^4].
\label{3.17}
\ee
Substituting this outcome into Eqs.(\ref{3.9},\ref{3.10}),
we are able to deduce that 
\be
\rho_r(0)=-\rho_e(0)={6\over \omega\pi G_4 M \epsilon^4} 
+{\cal O}\left[\epsilon^{-2}\right],
\label{3.18}
\ee
which will prove to be useful later in the paper.
\par 
We would not be so bold as to suggest that the above framework 
represents a physically realistic cosmology. On the
other hand, this scenario does have some desirable
and interesting features. For instance, the  four-dimensional cosmological
constant can be regarded as an input from the bulk 
theory and, therefore, the  {\it cosmological constant problem} \cite{WEI}
({\it i.e.}, how to explain  the relative smallness of $\Lambda_4$)  
may have a natural resolution
in terms of higher-dimensional dynamics. (This assumes
that string or M theory 
can be used to predict the values of $L$ and the brane
tension and that this prediction complies with the empirical
data. A very tall order indeed!) Furthermore, 
 the feature of a bouncing universe would allow one
to circumvent the issue of resolving the big bang (or crunch)
singularity, which  afflicts many (if not most) cosmological models.
\par
On the other hand,  the non-singular
 bounce depends on the existence of a negative-energy
matter source, $\rho_{e}$, which may be an unappealing feature to some.
It is, however, interesting to note that the {\it total}
energy density does  remain strictly
 non-negative\footnote{In spite of this feature of
non-negativity, the null energy condition (\ref{1.1})
 is still clearly violated, as can be seen by way of 
Eqs.(\ref{3.12},\ref{3.13},\ref{3.18}). Indeed, for
an open or flat FRW universe, such a violation is
a necessary pre-requisite for a non-singular bounce \cite{BLAH}.} 
and  the negative-energy source
will quickly  be  diluted away by  the cosmological expansion
(by virtue of the $a^{-6}$ red-shift factor). It is clear that, if 
one does not wish to call upon the rather dubious \cite{BV} null 
energy 
condition, some other means is necessary for 
testing the viability  of such exotic cosmologies.
In the next section, we will  invoke the causal energy
bound for just this purpose. 

\section{Causal Entropy Bound on the Brane}

As alluded to above, we now intend to employ the
causal entropy bound, as discussed in Section 2,
as a means for assessing the feasibility of
the exotic cosmologies in  Section 3.
To apply the causal bound, one should choose
 a given spacelike hypersurface and
then calculate  both the upper bound (\ref{2.9}) and the actual
entropy contained on that surface. Ideally,
every such surface in the spacetime should be tested;
however,  
in our study, it is sufficient to consider
the spacelike hypersurface at the bounce ($t=\eta=0$).
Significantly, this surface is precisely  where the  brane world 
will be most vulnerable to a violation of the bound.
\par
An important issue to resolve  is the means of
calculating the actual  entropy. For our model, there
are two entropic contributions to  consider: 
the radiative matter and 
the exotic stiff matter (the vacuum energy, of course, makes
no contribution). For simplicity,
we will only deal  with the former source, as one
might anticipate that radiation would make the  
dominant  contribution to the entropy. Moreover, if
we can verify that the bound is satisfied 
for the radiative entropy ($S_r$), the
inclusion of any additional  source only
strengthens our finding.
\par
To  be more explicit, we want to determine
if the inequality
\be
\left({S_{CB}\over S_r}\right)^2 >1
\label{4.1}
\ee
can be satisfied at the bounce, where 
\be
S_{CB} \equiv  \beta{VR_{CC}^{-1}\over G_4}
\label{4.2}
\ee
has been defined in accordance with Eq.(\ref{2.9}).
Let us also recall that the causal connection scale, $R_{CC}$,
can be obtained from either Eq.(\ref{2.5}) or Eq.(\ref{2.8}),
while $V$ denotes the volume of the hypersurface in question.
\par
As an initial step in this procedure,  $R_{CC}$ will be calculated by way
of Eq.(\ref{2.8}).  Including all three sources
(vacuum, radiation and exotic) and applying 
Eqs.(\ref{3.11}-\ref{3.13}) for
the corresponding pressures, we find that
\be
R_{CC}^{-2}={8\pi G_4\over 3}
{\rm Max}\left[2\rho_v-\rho_e,\quad 2\rho_r+3\rho_e\right].
\label{4.3}
\ee
Now focusing on the bounce surface in particular, we
see from Eq.(\ref{3.18}) that the left-hand argument
must be positive, whereas the right-hand argument is
clearly negative. Hence:
\be
R_{CC}^{-2}\approx {8\pi G_4\over 3} \rho_r, 
\label{4.4}
\ee
where we have made explicit use of  Eq.(\ref{3.18}), as well as
 the relative smallness
of the vacuum contribution. 
\par
 Let us next consider the  entropy of the radiation, $S_r$.
Here, it is helpful  to recall 
the standard Stephan-Boltzmann thermodynamic relations
for radiative matter (in thermal equilibrium at temperature $T$):
\be
s_r \equiv {S_r\over V} = {\cal N} \alpha_1 T^3,
\label{4.5}
\ee
\be
\rho_r ={\cal N} \alpha_2 T^4,
\label{4.6}
\ee
where the $\alpha$'s are numerical factors of
the order unity and ${\cal N}$ is the (effective) number of particle
species. Eliminating $T$ from this pair of equations,
we have
\be
S_r=V{\cal N}^{1/ 4}\alpha \rho_r^{3/ 4},
\label{4.7}
\ee
where $\alpha$ is another factor of the order unity. It
is somewhat unclear, in this case of
holographically induced radiation, as to
what  the actual value of ${\cal N}$ might be. However,
it is difficult to  imagine that ${\cal N}$ would be significantly
larger than $10^4$; therefore, we  will simply
absorb ${\cal N}^{1/4}$ into
the numerical factor, $\alpha$.
\par
Substituting Eqs.(\ref{4.2},\ref{4.4},\ref{4.7}) into
Eq.(\ref{4.1}) and simplifying, we now obtain
the following  inequality:
\be
{8\pi\over 3}{\beta^2\over \alpha^2}{\rho_r^{-1/2}\over G_4} >1.
\label{4.8}
\ee
We can make
this relation   more transparent by substituting for 
$\rho_r=\rho(0)$ ({\it cf}, Eq.(\ref{3.18})), 
\be
{8\pi\over 3}{\beta^2\over\alpha^2} 
\sqrt{{\omega\pi M \over 6 G_4}}\epsilon^2 >1,
\label{4.9}
\ee
and then dropping the inconsequential numerical factors to give
\be
\epsilon^2 > {1\over M_p}{1\over \sqrt{\omega M}},
\label{4.10}
\ee
where $M_p\sim G^{-1/2}_4$ is the Planck mass. 
\par
It is now clear that, for the casual entropy bound
to be protected, the parameter $|\epsilon|$
- or  the magnitude of the charge in units of 
black hole mass - must not become too small.
Although somewhat counter-intuitive, this outcome
makes sense if one considers that, in the proximity of the bounce,
the exotic energy density  goes as $\epsilon^{-4}$ 
({\it cf}, Eq.(\ref{3.18})),
rather than the naive expectation of $\epsilon^2$
({\it cf}, Eq.(\ref{3.10})). That is to say, a sufficiently
strong charge is needed to prevent the negative
energy from becoming too large in magnitude (at the bounce).
\par
The need for a lower bound on the charge  is certainly clear, but can
we get a better feel for the numbers involved?
First, let us consider the mass, $M$, of the bulk black hole.
From the brane-world perspective, $M$ should not be so large that
it endangers  early-universe cosmological
considerations, such as nucleosynthesis (although these kinds of 
constraints are highly model dependent \cite{RUB}). 
On the other hand, from the bulk point of view,
one would expect that the black hole mass is 
large enough  to insure stability. As is well known, an
anti-de Sitter Schwarzschild  black hole 
requires a radius  of at least $L$ 
to prevent a very rapid decay via Hawking radiation \cite{HAPA}.
The analagous statement here
(by an inspection of Eq.(\ref{3.2}))
 is essentially $\omega M > L^2$. 
Combining the bulk and brane  viewpoints, we can anticipate
that $\omega M\sim L^2$, and so (roughly speaking): 
\be 
\epsilon^2 > {L^{-1}\over M_p}.
\label{4.11}
\ee
\par
What about $L$?  The value of the bulk curvature
 in brane-world scenarios is a highly model-dependent feature \cite{RUB}:
 $L^{-1}$ is usually taken as being
anywhere from the electroweak scale (the lower limit
allowed by observation) up to  the four-dimensional
Planck scale itself. In all likelihood, the ``true'' value
would be somewhere in between; however, for the sake of argument,
let us consider the optimistic electroweak limit.
In this case:
\be
\epsilon^2 >  10^{-16};
\label{4.12}
\ee
meaning that the charge to mass ratio of the black hole
can still be negligibly small without jeopardizing
the causal entropy bound.

\section{Conclusion}

In summary,  we have ``re-opened the case'' on an intriguing
class of brane-world  scenarios. To elaborate, 
we have been investigating the validity, from a holographic perspective,
of  a  positively curved brane world
 evolving in the  background of a  Reissner-Nordstrom-like
black hole. These induced cosmologies are  particularly of interest 
for a couple of reasons. Firstly, 
as long as the bulk charge is non-vanishing, the brane world
is completely free of singularities.
That is, the brane universe  ``bounces'' when a  finite minimal
size is reached and thereby avoids the unpleasant implications
of a  big bang or crunch. Secondly, the holographically
induced matter includes an exotic, negative-energy 
 contribution. This exotic stiff matter arises
as a direct consequence of the electrostatic charge
in the bulk black hole, and it is clear that
the above two  features (bouncing universe and  negative-energy
source) are, in fact,  closely related.
\par
In a previous paper \cite{AJM}, we utilized 
a de Sitter-generalized $c$-theorem \cite{DS1,DS2,DS3,DS4} to argue that 
the same  class of cosmologies is not holographically 
viable. (For closely related works, see \cite{MCI,AJM3}.)
It is not, however, absolutely clear that
 the proposed $c$-theorem, which has 
 the  weak energy condition  as an
antecedent, can be  directly applied to
a spacetime that {\it a priori} violates this condition.  
It is also not apparent what a violation of
this $c$-theorem is precisely supposed to signify.
In this regard, it has been suggested by Strominger \cite{DS1} that 
the associated renormalization-group flow 
is dually related to the flow of time in the cosmology.
However,  the ambiguous nature of  time  in  quantum gravity \cite{ISH}
casts some doubt on the universal applicability of such an 
interpretation. Furthermore,
the $c$-theorem in question can be viewed as a manifestation
of the ``dS/CFT correspondence'' \cite{STR}, and, perhaps significantly,
this conjectured duality has been recently challenged 
on conceptual grounds \cite{SAF}.
\par
In view of such  interpretative difficulties,
we felt compelled to  re-examine the viability of these exotic cosmologies.
Again calling on holography, we have applied the causal entropy
bound  \cite{CAU}; specifically, 
at the most vulnerable point in the evolution of the brane world.
(This being at the bounce, where the negative-energy
matter plays its most dominant role.) From the perspective
of this holographic bound, there appears to be no
significant difficulties with the cosmologies of interest.
It has also been established  that the bulk charge must satisfy a lower bound;
however, this bound was  shown to be not particularly stringent for
a wide range of physically interesting scenarios.
\par
Unlike the priorly discussed $c$-theorem analysis, the causal
entropy bound has no interpretative difficulties. Rather,
this entropic bound follows from straightforward
arguments of causality and holography at its most
fundamental level. It is also significant that
the causal entropy bound does not employ 
the null (or weak) energy condition as an explicit premise.
 It is this feature, in particular,
that makes the causal bound an especially useful tool
 for discriminating spacetimes with
negative-energy sources.
\par
In spite of the above points, it is still quite possible
that the $c$-theorem violation should be taken seriously enough
to censor against these exotic cosmologies.  Nonetheless,
the current study casts significant doubt on
this viewpoint and signifies that, if nothing else,
the status of negative-energy matter (in a cosmological
framework) should remain a very open question.

\section{Acknowledgments}
\par
The author  would like to thank  V.P.  Frolov  for helpful
conversations.  
The author  also expresses his  gratitude
to M.E. Medved for her patience  
and encouragement.



\end{document}